\documentclass[twocolumn,showpacs,prb]{revtex4}
\usepackage{amssymb}
\usepackage{makeidx}
\usepackage{amsmath}
\usepackage{graphicx}
\usepackage{dcolumn}
\usepackage{bm}

\begin{document}

\title{Creation of particle-hole superposition states in graphene at
multiphoton resonant excitation by laser radiation}
\author{H.K. Avetissian}
\email{avetissian@ysu.am}
\author{A.K. Avetissian}
\author{G.F. Mkrtchian}
\author{Kh.V. Sedrakian}
\affiliation{Centre of Strong Fields Physics, Yerevan State University, 1 A. Manukian,
Yerevan 0025, Armenia }
\date{\today }

\begin{abstract}
Nonlinear dynamics of establishment of electron-hole coherent superpositions
states in graphene by multiphoton resonant excitation of interband
transitions in laser fields is considered. The single-particle time
dependent density matrix for such a quantized system is calculated in the
multiphoton resonant approximation. The dependence of Rabi oscillations of
Fermi-Dirac sea in graphene on the time, momentum, and photon number at
multiphoton laser-excitation is analyzed.
\end{abstract}

\date{\today }
\maketitle

\section{Introduction}

Graphene, a single sheet of carbon atoms in a honeycomb lattice has
attracted enormous interest since its experimental discovery and isolation.%
\cite{Nov1,Nov2} Its quasiparticle states behave like massless
\textquotedblleft relativistic\textquotedblright\ Dirac fermions \cite%
{Gaim,Nov3} and obey a two-dimensional Dirac equation, where the light speed
is replaced by the Fermi velocity that is 300 times smaller than speed of
light in vacuum. Beside various applications in electronic devices, graphene
physics opens wide research field unifying low-energy condensed matter
physics and quantum electrodynamics (QED).\cite%
{GQED1,GQED2,GQED4,GQED5,GQED6} Many fundamental nonlinear QED processes,
specifically, electron-positron pair production in superstrong laser fields
of ultrarelativistic intensities, \cite{Book} observation of which is
problematic yet even in the current superintense laser fields, have their
counterparts in graphene where considerably weaker electromagnetic fields
are required for experimental realization of \textquotedblleft antimatter\textquotedblright\ production from
vacuum. In this connection one can note Klein paradox, \cite{K1,K2,K3}
Schwinger mechanism \cite{S1,S2,S3} and Zitterbewegung \cite{Z1,Z2,Z3} for
particle-hole excitation, as well as diverse physical and applied effects
based on Zitterbewegung, e.g., minimal conductivity at vanishing carrier
concentration \cite{cn1,cn2} etc. \ \ \ \ 

Due to massless energy spectrum the Compton wavelength for graphene
quasiparticle tends to infinity. On the other hand, in QED the Compton
wavelength is characteristic length for particle-antiparticle pair creation
and annihilation. So, at the interaction of an electromagnetic field with an
intrinsic graphene there is not a quasiclassical limit, since no matter how
weak the applied field is and how small the photon energy is, the
particle-hole pairs will be created during the whole interaction process -at
arbitrary distances.

In graphene wave-particle interaction can be characterized by the
dimensionless parameter%
\begin{equation*}
\chi =\frac{eE\mathrm{v}_{F}}{\omega }\frac{1}{\hbar \omega },
\end{equation*}%
which represents the work of the wave electric field $E$ on a period $%
1/\omega $ in the units of photon energy $\varepsilon _{\gamma }=\hbar
\omega $. Here $\mathrm{v}_{F}$ is the Fermi velocity: $\mathrm{v}%
_{F}\approx c/300$,$\ $and $e$ is the elementary charge. The average
intensity of the wave expressed by $\chi $, can be estimated as:%
\begin{equation*}
I_{\chi }=\chi ^{2}\times 3.07\times 10^{11}\ \mathrm{W\ cm}^{-2}[\hbar
\omega /\mathrm{eV}]^{4}.
\end{equation*}%
Depending on the value of this parameter $\chi $,\ one can distinguish three
different regimes in the wave-particle interaction process. Thus,\ $\chi <<1$
corresponds to one-photon interaction regime, $\chi \sim 1$ -- to
multiphoton interaction regime, and $\chi >>1$ corresponds to static field
limit or Schwinger regime. As is seen, the intensity $I_{\chi }$ strongly
depends on the photon energy. Particularly, for infrared photons: $%
\varepsilon _{\gamma }\sim 0.1$ $\mathrm{eV}$, multiphoton interaction
regime can be achieved at the intensities $I_{\chi }=3.07\times 10^{7}\ 
\mathrm{W\ cm}^{-2}$. Note that in case of free electrons at the same photon
energies multiphoton effects take place at the intensities $I\sim 10^{16}\ 
\mathrm{W\ cm}^{-2}$. \cite{Book} Such a huge difference, as well as the
gapless particle-hole energy spectrum in graphene makes realistic another
interesting nonlinear QED process: multiphoton excitation of Dirac vacuum
with Rabi oscillations at ultrafast excitation. \cite{pair}

In the present work the creation of particle-hole coupled states in graphene
via multiphoton resonant excitation by laser fields is studied. It is well
known that Rabi oscillation of states' populations is the coherent response
of two-level atomic systems under resonant excitation. The one-photon
resonant excitation of atoms and associated Rabi oscillations have been
comprehensively studied both theoretically and experimentally and described
in numerous review articles and books (see, e.g., \cite{Eb-Sh}). Similar
phenomena have also been observed in semiconductors.\cite{Rs1} Recently, the
Rabi oscillations in graphene\ at one-photon interband excitation (at $\chi
<<1$) and its influence on the dynamic conductivity were investigated in
Refs. [\onlinecite{Rabi1,Rabi2,Rabi3}]. On the other hand, at a
wave-particle interaction in graphene due to free-free intraband transitions
we have a situation analogous to resonant excitation of quantum systems with
permanent dipole moments, where direct multiphoton transitions are very
effective.\cite{MRE1,MRE2,MRE3} Hence, it is of interest the creation of
particle-hole coupled states in graphene via multiphoton resonant
excitation. This process apart from fundamental interest may also have
practical applications. In particular, particle-hole annihilation from
coherent superposition states will cause intense coherent radiation of
harmonics of the applied wave-field. We consider multiphoton interaction
regime at $\chi \sim 1$. Accordingly, the time evolution of considered
process is found using a nonperturbative resonant approach arising from the
quantum kinetic equations.

The paper is organized as follows. In Sec. II the set of equations for
single-particle density matrix is formulated. In Sec. III we present the
solution of those equations in the multiphoton resonant approximation. In
Sec. IV the results of corresponding numerical simulations are presented.
Finally, conclusions are given in Sec. V.

\section{Basic model and evolutionary equation for single-particle density
matrix}

Let graphene interacts with a plane quasimonochromatic laser radiation of
carrier frequency $\omega $ and slowly varying envelope. To clarify the
picture of stated in this paper problem for creation of coherent
superposition states, we consider the case of interaction when the laser
wave propagates in perpendicular direction to graphene plane ($XY$) to
exclude the effect of magnetic field. This travelling wave for electrons in
graphene becomes a homogeneous time-periodic electric field. Let it to be
directed along the $X$-axis with the form (constant phase connected with the
position of the wave pulse maximum with respect to graphene plane is set
zero): 
\begin{equation}
\mathbf{E}\left( t\right) =\widehat{x}E_{0}\cos \omega t.  \label{Et}
\end{equation}%
The problem that we attempt to solve in the given field approximation is
analogous to Rabi oscillations in two level atomic systems with permanent
dipole moments. In this case, the physical picture of resonant wave-graphene
interaction will be more transparent in the length gauge. So, for the
interaction Hamiltonian we will use length gauge describing the interaction
by the potential energy. Cast in the second quantization formalism, the
Hamiltonian is%
\begin{equation}
\widehat{H}=\int \widehat{\Psi }^{+}\widehat{H}_{s}\widehat{\Psi }dxdy,
\label{ham}
\end{equation}%
where $\widehat{\Psi }$ is the fermionic field operator, $\widehat{H}_{s}$
is the single-particle Hamiltonian in the external homogeneous electric
field (\ref{Et}). Omitting here real spin and valley quantum numbers,
single-particle Hamiltonian can be written as:%
\begin{equation}
\widehat{H}_{s}=\mathrm{v}_{F}\left( 
\begin{array}{cc}
0 & \widehat{p}_{x}-i\widehat{p}_{y} \\ 
\widehat{p}_{x}+i\widehat{p}_{y} & 0%
\end{array}%
\right) +\left( 
\begin{array}{cc}
exE & 0 \\ 
0 & exE%
\end{array}%
\right) ,  \label{sham}
\end{equation}%
where $\mathrm{v}_{F}\approx c/300$ is the Fermi velocity ($c$ is the light
speed in vacuum),\textbf{\ }$\mathbf{\hat{p}}=\left\{ \widehat{p}_{x},%
\widehat{p}_{y}\right\} $\textbf{\ }is the electron momentum operator. The
first term in (\ref{sham}) is the Hamiltonian of two-dimensional massless
Dirac fermion and the second term is the interaction part.

We write the fermionic field operator in the form of an expansion in the
free Dirac states:%
\begin{equation}
\widehat{\Psi }(x,y,t)=\sum\limits_{\mathbf{p,}\sigma }\widehat{a}_{\mathbf{p%
},\sigma }(t)\Psi _{\mathbf{p,}\sigma }(x,y),  \label{exp}
\end{equation}%
where the creation and annihilation operators, $\widehat{a}_{\mathbf{p}%
,\sigma }^{+}(t)$ and $\widehat{a}_{\mathbf{p},\sigma }(t)$, associated with
positive ($\sigma =1$) and negative ($\sigma =-1$) energy solutions satisfy
the anticommutation rules at equal times%
\begin{equation}
\{\widehat{a}_{\mathbf{p},\sigma }^{\dagger }(t),\widehat{a}_{\mathbf{p}%
^{\prime },\sigma ^{\prime }}(t^{\prime })\}_{t=t^{\prime }}=\delta _{%
\mathbf{p,p^{\prime }}}\delta _{\sigma ,\sigma ^{\prime }}\;,\;  \label{com1}
\end{equation}%
\begin{equation}
\{\widehat{a}_{\mathbf{p},\sigma }^{\dagger }(t),\widehat{a}_{\mathbf{p}%
^{\prime },\sigma ^{\prime }}^{\dagger }(t^{\prime })\}_{t=t^{\prime }}=\{%
\widehat{a}_{\mathbf{p},\sigma }(t),\widehat{a}_{\mathbf{p}^{\prime },\sigma
^{\prime }}(t^{\prime })\}_{t=t^{\prime }}=0.  \label{com2}
\end{equation}%
The free Dirac solutions corresponding to energies $\mathcal{E}_{\sigma
}=\sigma \mathrm{v}_{F}\sqrt{p_{x}^{2}+p_{y}^{2}}\ $($\sigma =\pm 1$) are:%
\begin{equation}
\Psi _{\mathbf{p,}\sigma }(x,y)=\frac{1}{\sqrt{2S}}\left( 
\begin{array}{c}
1 \\ 
\sigma e^{i\Theta \left( \mathbf{p}\right) }%
\end{array}%
\right) e^{\frac{i}{\hbar }\left( p_{x}x+p_{y}y\right) },  \label{free}
\end{equation}%
where $S$ is the quantization area (graphene layer surface area) and 
\begin{equation}
\Theta \left( \mathbf{p}\right) =\arctan \left( \frac{p_{y}}{p_{x}}\right)
\label{angle}
\end{equation}%
is the angle in momentum space.

Taking into account Eqs. (\ref{Et})-(\ref{angle}), the second quantized
Hamiltonian can be expressed in the form:%
\begin{equation*}
\widehat{H}=\sum\limits_{\mathbf{p,}\sigma }\mathcal{E}_{\sigma }\left(
p\right) \widehat{a}_{\mathbf{p}\sigma }^{+}\widehat{a}_{\mathbf{p}\sigma }
\end{equation*}%
\begin{equation}
+eE\left( t\right) \sum\limits_{\mathbf{p,}\sigma }\sum\limits_{\mathbf{p}%
^{\prime },\sigma ^{\prime }}D_{\sigma \sigma ^{\prime }}\left( \mathbf{p,p}%
^{\prime }\right) \widehat{a}_{\mathbf{p,}\sigma }^{+}\widehat{a}_{\mathbf{p}%
^{\prime },\sigma ^{\prime }},  \label{Ham}
\end{equation}%
where%
\begin{equation*}
D_{\sigma \sigma ^{\prime }}\left( \mathbf{p,p}^{\prime }\right) =\frac{1}{2S%
}\left[ 1+\sigma \sigma ^{\prime }e^{i\left( \Theta \left( \mathbf{p}%
^{\prime }\right) -\Theta \left( \mathbf{p}\right) \right) }\right] 
\end{equation*}%
\begin{equation}
\times \int xe^{\frac{i}{\hbar }\left( \mathbf{p}^{\prime }-\mathbf{p}%
\right) \mathbf{r}}dxdy.  \label{D}
\end{equation}%
We will use Heisenberg representation, where operators evolution are given
by the following equation:%
\begin{equation}
i\hbar \frac{\partial \widehat{L}}{\partial t}=\left[ \widehat{L},\widehat{H}%
\right] ,  \label{Heis}
\end{equation}%
and expectation values are determined by the initial density matrix $%
\widehat{D}$:%
\begin{equation}
<\widehat{L}>=\mathrm{Sp}\left( \widehat{D}\widehat{L}\right) .  \label{aver}
\end{equation}%
The single-particle density matrix in momentum space is defined as:%
\begin{equation}
\rho _{\sigma _{1}\sigma _{2}}(\mathbf{p}_{1},\mathbf{p}_{2},t)=<\widehat{a}%
_{\mathbf{p}_{2},\sigma _{2}}^{+}(t)\widehat{a}_{\mathbf{p}_{1},\sigma
_{1}}(t)>.  \label{SPDM}
\end{equation}%
For initial state of graphene quasiparticles we assume ideal Fermi gas in
equilibrium. This means that the initial single-particle density matrix is
diagonal and we have Fermi-Dirac distribution:%
\begin{equation}
\rho _{\sigma \sigma ^{\prime }}(\mathbf{p},\mathbf{p}^{\prime },0)=\frac{1}{%
1+e^{\frac{\mathcal{E}_{\sigma }\left( p\right) -\mu }{T}}}\delta _{\mathbf{%
p,p^{\prime }}}\delta _{\sigma ,\sigma ^{\prime }}.  \label{ISPDM}
\end{equation}%
Here $\mu $ is the chemical potential, $T$ is the temperature in energy
units. Taking into account definition (\ref{SPDM}), from Eq. (\ref{Heis})
one can obtain evolution equation for single-particle density matrix: 
\begin{equation*}
i\hbar \frac{\partial \rho _{\sigma _{1}\sigma _{2}}(\mathbf{p}_{1},\mathbf{p%
}_{2},t)}{\partial t}=\left[ \mathcal{E}_{\sigma _{1}}\left( p_{1}\right) -%
\mathcal{E}_{\sigma _{2}}\left( p_{2}\right) \right] \rho _{\sigma
_{1}\sigma _{2}}(\mathbf{p}_{1},\mathbf{p}_{2},t)
\end{equation*}%
\begin{equation*}
-eE\left( t\right) \sum\limits_{\mathbf{p,}\sigma }\left[ D_{\sigma \sigma
_{2}}\left( \mathbf{p,p}_{2}\right) \rho _{\sigma _{1}\sigma }(\mathbf{p}%
_{1},\mathbf{p},t)\right. 
\end{equation*}%
\begin{equation}
\left. -D_{\sigma _{1}\sigma }\left( \mathbf{p}_{1}\mathbf{,p}\right) \rho
_{\sigma \sigma _{2}}(\mathbf{p},\mathbf{p}_{2},t)\right] .  \label{evol}
\end{equation}%
Then taking into account the following relation with Dirac delta function $%
\delta \left( \alpha \right) $: 
\begin{equation*}
\int_{-\infty }^{\infty }xe^{-i\alpha x}dx=2\pi i\frac{\partial }{\partial
\alpha }\delta \left( \alpha \right) 
\end{equation*}%
and substitution 
\begin{equation*}
\sum\limits_{\mathbf{p}}\rightarrow \frac{S}{(2\pi \hbar )^{2}}\int d\mathbf{%
p,}
\end{equation*}%
we obtain closed set of equation for the density matrix elements:%
\begin{equation*}
\frac{\partial \rho _{\sigma ,\sigma }(\mathbf{p},\mathbf{p},t)}{\partial t}%
-eE\left( t\right) \left. \frac{\partial \rho _{\sigma ,\sigma }(\mathbf{p}%
_{1},\mathbf{p},t)}{\partial p_{1x}}\right\vert _{\mathbf{p}_{1}=\mathbf{p}}
\end{equation*}%
\begin{equation*}
-eE\left( t\right) \left. \frac{\partial \rho _{\sigma ,\sigma }(\mathbf{p},%
\mathbf{p}_{2},t)}{\partial p_{2x}}\right\vert _{\mathbf{p}_{2}=\mathbf{p}}=i%
\frac{eE\left( t\right) }{2}\frac{\partial \Theta \left( \mathbf{p}\right) }{%
\partial p_{x}}
\end{equation*}%
\begin{equation}
\times \left[ \rho _{\sigma ,-\sigma }(\mathbf{p},\mathbf{p},t)-\rho
_{-\sigma ,\sigma }(\mathbf{p},\mathbf{p},t)\right] ,  \label{R_ss}
\end{equation}%
\begin{equation*}
\frac{\partial \rho _{\sigma ,-\sigma }(\mathbf{p},\mathbf{p},t)}{\partial t}%
-eE\left( t\right) \left. \frac{\partial \rho _{\sigma ,-\sigma }(\mathbf{p}%
_{1},\mathbf{p},t)}{\partial p_{1x}}\right\vert _{\mathbf{p}_{1}=\mathbf{p}}
\end{equation*}%
\begin{equation*}
-eE\left( t\right) \left. \frac{\partial \rho _{\sigma ,-\sigma }(\mathbf{p},%
\mathbf{p}_{2},t)}{\partial p_{2x}}\right\vert _{\mathbf{p}_{2}=\mathbf{p}}=%
\frac{2}{i\hbar }\mathcal{E}_{\sigma }\left( p\right) \rho _{\sigma ,-\sigma
}(\mathbf{p},\mathbf{p},t)
\end{equation*}%
\begin{equation}
-\frac{eE\left( t\right) }{2i}\frac{\partial \Theta \left( \mathbf{p}\right) 
}{\partial p_{x}}\left[ \rho _{\sigma ,\sigma }(\mathbf{p},\mathbf{p}%
,t)-\rho _{-\sigma ,-\sigma }(\mathbf{p},\mathbf{p},t)\right] .
\label{R_s_s}
\end{equation}

In Eqs. (\ref{R_ss}) and (\ref{R_s_s}) one can eliminate the terms with
partial derivatives $\partial /\partial p_{x}$ by the method of
characteristics. The characteristic of these equations is the classical
equation of motion: 
\begin{equation*}
\frac{d\mathbf{p}}{dt}=-e\mathbf{E}\left( t\right)
\end{equation*}%
with the solution 
\begin{equation}
p_{x}=p_{0x}+p_{E}\left( t\right) ;\ p_{y}=p_{0y},  \label{sol}
\end{equation}%
where 
\begin{equation*}
p_{E}\left( t\right) =-e\int_{0}^{t}E\left( t^{\prime }\right) dt^{\prime }=-%
\frac{eE_{0}}{\omega }\sin \omega t
\end{equation*}%
is the momentum given by the wave-field. Thus, with the new variables $%
p_{0x} $, $p_{0y}$, and $t$ Eqs. (\ref{R_ss}) and (\ref{R_s_s}) read:%
\begin{equation*}
\frac{\partial \rho _{\sigma ,\sigma }(\mathbf{p}_{0},\mathbf{p}_{0},t)}{%
\partial t}=\frac{i}{2}F\left( \mathbf{p}_{0},t\right)
\end{equation*}%
\begin{equation}
\times \left[ \rho _{\sigma ,-\sigma }(\mathbf{p}_{0},\mathbf{p}_{0},t)-\rho
_{-\sigma ,\sigma }(\mathbf{p}_{0},\mathbf{p}_{0},t)\right] ,  \label{R_ss1}
\end{equation}%
\begin{equation*}
\frac{\partial \rho _{\sigma ,-\sigma }(\mathbf{p}_{0},\mathbf{p}_{0},t)}{%
\partial t}=\frac{2}{i\hbar }\widetilde{\mathcal{E}}_{\sigma }\left( \mathbf{%
p}_{0},t\right) \rho _{\sigma ,-\sigma }(\mathbf{p}_{0},\mathbf{p}_{0},t)
\end{equation*}%
\begin{equation}
+\frac{i}{2}F\left( \mathbf{p}_{0},t\right) \left[ \rho _{\sigma ,\sigma }(%
\mathbf{p}_{0},\mathbf{p}_{0},t)-\rho _{-\sigma ,-\sigma }(\mathbf{p}_{0},%
\mathbf{p}_{0},t)\right] ,  \label{R_s_s1}
\end{equation}%
where 
\begin{equation}
F\left( \mathbf{p}_{0},t\right) =-\frac{eE\left( t\right) p_{0y}}{\left(
p_{0x}+p_{E}\left( t\right) \right) ^{2}+p_{0y}^{2}}  \label{F}
\end{equation}%
and%
\begin{equation}
\widetilde{\mathcal{E}}_{\sigma }\left( \mathbf{p}_{0},t\right) =\sigma 
\mathrm{v}_{F}\sqrt{\left( p_{0x}+p_{E}\left( t\right) \right)
^{2}+p_{0y}^{2}}.  \label{Egt}
\end{equation}%
Taking into account Eq. (\ref{sol}), it is easy to see that

\begin{equation}
\rho _{\sigma ,\sigma ^{\prime }}(\mathbf{p}_{0},\mathbf{p}_{0},0)=\rho
_{\sigma ,\sigma ^{\prime }}(\mathbf{p},\mathbf{p},0).  \label{init}
\end{equation}

To be more precise in the set of equations (\ref{R_ss}) and (\ref{R_s_s})
one should add the terms describing relaxation processes. Since we have not
taken into account the relaxation processes, this consideration is correct
only for the times $t<\tau _{\min }$, where $\tau _{\min }$ is the minimum
of all relaxation times. Thus, full dynamics in the absence of any losses is
now governed by Eqs. (\ref{R_ss1}) and (\ref{R_s_s1}). These equations yield
the conservation law for the particle number:%
\begin{equation*}
\rho _{1,1}(\mathbf{p}_{0},\mathbf{p}_{0},t)+\rho _{-1,-1}(\mathbf{p}_{0},%
\mathbf{p}_{0},t)=
\end{equation*}%
\begin{equation}
=\rho _{1,1}(\mathbf{p}_{0},\mathbf{p}_{0},0)+\rho _{-1,-1}(\mathbf{p}_{0},%
\mathbf{p}_{0},0)\equiv \Xi \left( p_{0},\mu ,T\right) .  \label{cons}
\end{equation}%
Here we have introduced the notation $\Xi _{p_{0},\mu ,T}$, which according
to Eq. (\ref{ISPDM}) is: 
\begin{equation*}
\Xi _{p_{0},\mu ,T}=\frac{1}{1+e^{\frac{\mathrm{v}_{F}p_{0}-\mu }{T}}}+\frac{%
1}{1+e^{\frac{-\mathrm{v}_{F}p_{0}-\mu }{T}}}.
\end{equation*}%
In Eqs.(\ref{R_ss1}) and (\ref{R_s_s1}) diagonal elements represent particle 
$\mathcal{N}\left( \mathbf{p}_{0},t\right) \equiv \rho _{1,1}(\mathbf{p}_{0},%
\mathbf{p}_{0},t)$ and hole ($1-\rho _{-1,-1}(\mathbf{p}_{0},\mathbf{p}%
_{0},t)$) distribution functions, while nondiagonal elements $\rho _{1,-1}(%
\mathbf{p}_{0},\mathbf{p}_{0},t)=\rho _{-1,1}^{\ast }(\mathbf{p}_{0},\mathbf{%
p}_{0},t)$ describe particle-hole coherent transitions. Introducing the new
quantity ($\mathcal{J}\left( \mathbf{p}_{0},t\right) $):%
\begin{equation*}
\rho _{1,-1}(\mathbf{p}_{0},\mathbf{p}_{0},t)=i\mathcal{J}\left( \mathbf{p}%
_{0},t\right) \exp \left\{ -i\frac{2}{\hbar }\int_{0}^{t}\widetilde{\mathcal{%
E}}_{1}\left( \mathbf{p}_{0},t^{\prime }\right) dt^{\prime }\right\}
\end{equation*}%
and taking into account that $\rho _{-1,-1}(\mathbf{p}_{0},\mathbf{p}%
_{0},t)=\Xi _{p_{0},\mu ,T}-\mathcal{N}\left( \mathbf{p}_{0},t\right) $,
from Eqs.(\ref{R_ss1}) and (\ref{R_s_s1}) we obtain the following set of
equations:%
\begin{equation*}
\frac{\partial \mathcal{N}\left( \mathbf{p}_{0},t\right) }{\partial t}=-%
\frac{1}{2}F\left( \mathbf{p}_{0},t\right)
\end{equation*}%
\begin{equation}
\times \left[ \mathcal{J}\left( \mathbf{p}_{0},t\right) \exp \left\{ -i\frac{%
2}{\hbar }\int_{0}^{t}\widetilde{\mathcal{E}}_{1}\left( \mathbf{p}%
_{0},t^{\prime }\right) dt^{\prime }\right\} +\mathrm{c.c.}\right] ,
\label{f1}
\end{equation}%
\begin{equation*}
\frac{\partial \mathcal{J}\left( \mathbf{p}_{0},t\right) }{\partial t}=\frac{%
1}{2}F\left( \mathbf{p}_{0},t\right) \exp \left\{ i\frac{2}{\hbar }%
\int_{0}^{t}\widetilde{\mathcal{E}}_{1}\left( \mathbf{p}_{0},t^{\prime
}\right) dt^{\prime }\right\}
\end{equation*}%
\begin{equation}
\times \left[ 2\mathcal{N}\left( \mathbf{p}_{0},t\right) -\Xi _{p_{0},\mu ,T}%
\right] .  \label{f2}
\end{equation}%
This set of equations should be solved with the initial conditions:

\begin{equation}
\mathcal{J}\left( \mathbf{p}_{0},0\right) =0;\quad \mathcal{N}\left( \mathbf{%
p}_{0},0\right) =\frac{1}{1+e^{\frac{\mathrm{v}_{F}p_{0}-\mu }{T}}}.
\label{Koshi}
\end{equation}

\section{Multiphoton resonant excitation}

The Eqs. (\ref{f1}) and (\ref{f2}) represent linear set of equations with
periodic coefficients and are analogous to Bloch equations, \cite{Eb-Sh}
which describe Rabi oscillation of states' populations of two-level atomic
system under resonant excitation. Note that there are significant
differences between usual Bloch equations and Eqs. (\ref{f1}), (\ref{f2}).
Thus, from the Floquet theorem and Eq. (\ref{Egt}) follows that instead of
stationary levels $\mathrm{v}_{F}p_{0}$\ and $-\mathrm{v}_{F}p_{0}$ due to
free-free intraband transitions we have quasistationary\ states with
quasienergies: $\widetilde{\mathcal{E}}_{\pm }\left( s\right) =\pm \mathcal{E%
}_{E_{0}}\left( \mathbf{p}_{0}\right) +s\omega ;$\ $s=0,\pm 1,\pm 2,...$.
Here 
\begin{equation*}
\mathcal{E}_{E_{0}}\left( \mathbf{p}_{0}\right) =\frac{\omega }{2\pi }%
\int_{0}^{2\pi /\omega }\widetilde{\mathcal{E}}_{1}\left( \mathbf{p}%
_{0},t\right) dt
\end{equation*}%
\begin{equation}
=\mathrm{v}_{F}\frac{\omega }{2\pi }\int_{0}^{2\pi /\omega }\sqrt{\left(
p_{0x}-\frac{eE_{0}}{\omega }\sin \omega t\right) ^{2}+p_{0y}^{2}}dt
\label{quasi}
\end{equation}%
is the mean value of classical energy in the field (\ref{Et}), which has a
nonlinear dependence on the amplitude of the wave-field. The latter in the
physical sense is the dynamic Stark shift due to free-free intraband
transitions. Two Floquet ladders are coupled by the term $F\left( \mathbf{p}%
_{0},t+2\pi /\omega \right) =F\left( \mathbf{p}_{0},t\right) $, which in
tern contains all harmonics of driving field. In usual Bloch equations
coupling contains only fundamental oscillations, which provide only direct
one photon resonant excitation. Situation is analogous to resonant
excitation of systems with permanent dipole moments, where, as has been
shown in Ref. [\onlinecite{MRE1}], it is possible to decouple slow and rapid
oscillations and to disclose the resonant dynamics of the wave-particle
interaction.

Because of space homogeneity of the field (\ref{Et}), the generalized
momentum of a particle conserves, so that the real transitions in the field
occur from a $-\mathcal{E}_{E_{0}}\left( \mathbf{p}_{0}\right) $ negative
energy level to the positive $+\mathcal{E}_{E_{0}}\left( \mathbf{p}%
_{0}\right) $ energy level and, consequently, the multiphoton probabilities
of particle-hole pair production will have maximal values for the resonant
transitions 
\begin{equation}
2\mathcal{E}_{E_{0}}\simeq n\hbar \omega .  \label{res}
\end{equation}%
Note that the resonant condition (\ref{res}) is equivalent to crossing of
Floquet ladders: $\widetilde{\mathcal{E}}_{-}\left( s+n\right) \simeq 
\widetilde{\mathcal{E}}_{+}\left( s\right) $. Figure 1 schematically
illustrates multiphoton interband transition between the two states in the
filled lower cone and the empty part of upper cone. 
\begin{figure}[tbp]
\includegraphics[width=.2\textwidth]{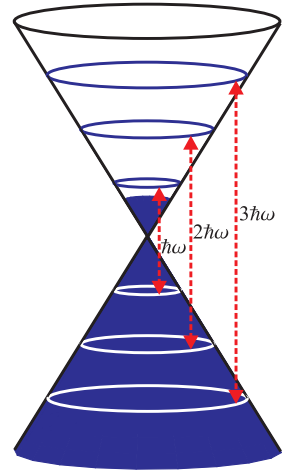}
\caption{Graphene conical dispersion with interband multiphoton transitions
induced by external electric field.}
\end{figure}
To decouple slow and rapid oscillations at the resonant condition (\ref{res}%
), we follow the ansatz developed in Ref. [\onlinecite{MRE1}] and represent
Eqs. (\ref{f1}) and (\ref{f2}) in the form: 
\begin{equation*}
\frac{\partial \mathcal{N}\left( \mathbf{p}_{0},t\right) }{\partial t}=-%
\frac{1}{2}\mathcal{J}\left( \mathbf{p}_{0},t\right) 
\end{equation*}%
\begin{equation}
\times \sum_{s}G_{s}^{\ast }\left( \mathbf{p}_{0},E_{0}\right) e^{-i\left( 
\frac{2}{\hbar }\mathcal{E}_{E}\left( \mathbf{p}_{0}\right) -s\omega \right)
t}+\mathrm{c.c.},  \label{ff1}
\end{equation}%
\begin{equation*}
\frac{\partial \mathcal{J}\left( \mathbf{p}_{0},t\right) }{\partial t}=\frac{%
1}{2}\sum_{s}G_{s}\left( \mathbf{p}_{0},E_{0}\right) e^{i\left( \frac{2}{%
\hbar }\mathcal{E}_{E}\left( \mathbf{p}_{0}\right) -s\omega \right) t}
\end{equation*}%
\begin{equation}
\times \left[ 2\mathcal{N}\left( \mathbf{p}_{0},t\right) -\Xi _{p_{0},\mu ,T}%
\right] .  \label{ff2}
\end{equation}%
Here $s$ photon coupling coefficient is: 
\begin{equation*}
G_{s}\left( \mathbf{p}_{0},E_{0}\right) =\frac{\omega }{2\pi }\int_{0}^{2\pi
/\omega }F\left( \mathbf{p}_{0},t\right) 
\end{equation*}%
\begin{equation}
\times \exp \left\{ -\sum_{m\neq 0}\frac{2\mathcal{E}_{m}\left( \mathbf{p}%
_{0}\right) }{m\hbar \omega }e^{-im\omega t}\right\} e^{is\omega t}dt,
\label{scoup}
\end{equation}%
where%
\begin{equation}
\mathcal{E}_{m}\left( \mathbf{p}_{0}\right) =\frac{\omega }{2\pi }%
\int_{0}^{2\pi /\omega }\widetilde{\mathcal{E}}_{1}\left( \mathbf{p}%
_{0},t\right) e^{im\omega t}dt.  \label{en_m}
\end{equation}%
Close to resonance (\ref{res}), the main coupling term in Eqs. (\ref{ff1})
and (\ref{ff2}) becomes slowly varying term with $s=n$. The remaining
nonresonant and rapidly oscillating terms are responsible only for dynamic
Stark shifts.\cite{MRE1} Thus, for time average functions $\overline{%
\mathcal{N}}\left( \mathbf{p}_{0},t\right) $ and $\overline{\mathcal{J}}%
\left( \mathbf{p}_{0},t\right) $ one can obtain the following set of
equations:%
\begin{equation}
\frac{\partial \overline{\mathcal{N}}\left( \mathbf{p}_{0},t\right) }{%
\partial t}=-\frac{1}{2}\overline{\mathcal{J}}\left( \mathbf{p}_{0},t\right)
G_{n}^{\ast }\left( \mathbf{p}_{0},E_{0}\right) e^{-i\delta _{n}t}+\mathrm{%
c.c.},  \label{fff1}
\end{equation}%
\begin{equation*}
\frac{\partial \overline{\mathcal{J}}\left( \mathbf{p}_{0},t\right) }{%
\partial t}+i\delta _{st}\overline{\mathcal{J}}\left( \mathbf{p}%
_{0},t\right) 
\end{equation*}%
\begin{equation}
=\frac{1}{2}G_{n}\left( \mathbf{p}_{0},E_{0}\right) e^{i\delta _{n}t}\left( 2%
\overline{\mathcal{N}}\left( \mathbf{p}_{0},t\right) -\Xi _{p_{0},\mu
,T}\right) .  \label{fff2}
\end{equation}%
Here we have introduced resonance detuning%
\begin{equation}
\delta _{n}=\frac{2\mathcal{E}_{E}\left( \mathbf{p}_{0}\right) -n\hbar
\omega }{\hbar }  \label{det}
\end{equation}%
and dynamic Stark shift%
\begin{equation}
\delta _{st}=\frac{1}{2\omega }\sum_{s\neq n}\frac{\left\vert G_{s}\left( 
\mathbf{p}_{0},E_{0}\right) \right\vert ^{2}}{\left( n-s\right) }.
\label{st}
\end{equation}%
The latter is the result of virtual nonresonant transition within Floquet
states. Thus, we have a set of linear ordinary differential equations the
solution of which at the initial condition (\ref{Koshi}) is:%
\begin{equation*}
\overline{\mathcal{J}}\left( \mathbf{p}_{0},t\right) =e^{i\delta _{n}t}\frac{%
G_{n}\left( \mathbf{p}_{0},E_{0}\right) }{2\Omega _{n}}\Delta _{p_{0},\mu ,T}
\end{equation*}%
\begin{equation}
\times \left( \sin \Omega _{n}t-i\frac{\delta _{n}+\delta _{st}}{\Omega _{n}}%
\left( 1-\cos \Omega _{n}t\right) \right) ,  \label{sol1}
\end{equation}%
\begin{equation*}
\overline{\mathcal{N}}\left( \mathbf{p}_{0},t\right) =\frac{\Xi _{p_{0},\mu
,T}}{2}+\frac{\left\vert G_{n}\left( \mathbf{p}_{0},E_{0}\right) \right\vert
^{2}}{2\Omega _{n}^{2}}\Delta _{p_{0},\mu ,T}
\end{equation*}%
\begin{equation}
\times \left[ \frac{\left( \delta _{n}+\delta _{st}\right) ^{2}}{\left\vert
G_{n}\left( \mathbf{p}_{0},E_{0}\right) \right\vert ^{2}}+\cos \Omega _{n}t%
\right] ,  \label{sol2}
\end{equation}%
where%
\begin{equation}
\Delta _{p_{0},\mu ,T}=\frac{1}{1+e^{\frac{\mathrm{v}_{F}p_{0}-\mu }{T}}}-%
\frac{1}{1+e^{\frac{-\mathrm{v}_{F}p_{0}-\mu }{T}}}  \label{inversion}
\end{equation}%
is the initial population inversion, and

\begin{equation}
\Omega _{n}=\sqrt{\left\vert G_{n}\left( \mathbf{p}_{0},E_{0}\right)
\right\vert ^{2}+\left( \delta _{n}+\delta _{st}\right) ^{2}}  \label{Rabi}
\end{equation}%
is the generalized Rabi frequency. The solution (\ref{sol2}) expresses Rabi
flopping among the particle-hole states at the multiphoton resonance. The
solutions (\ref{sol1}), (\ref{sol2}) have been derived using the assumption
that $\overline{\mathcal{N}}\left( \mathbf{p}_{0},t\right) $ and $\overline{%
\mathcal{J}}\left( \mathbf{p}_{0},t\right) $ are slowly varying functions on
the scale of the wave period, which put the following restrictions:

\begin{equation}
(\left\vert G_{n}\left( \mathbf{p}_{0},E_{0}\right) \right\vert ,\
\left\vert \delta _{n}\right\vert ,\ \left\vert \delta _{st}\right\vert
)<<\omega  \label{condition}
\end{equation}%
on the characteristic parameters of the system considered.

For the exact resonance ($\delta _{n}+\delta _{st}=0$) we have $\Omega
_{n}=\left\vert G_{n}\left( \mathbf{p}_{0},E_{0}\right) \right\vert $ and
the solutions become: 
\begin{equation}
\mathcal{J}\left( \mathbf{p}_{0},t\right) =\frac{\Delta _{p_{0},\mu ,T}}{2}%
e^{i\arg \left( G_{n}\left( \mathbf{p}_{0},E_{0}\right) \right) }\sin \Omega
_{n}t,  \label{sol3}
\end{equation}%
\begin{equation}
\mathcal{N}\left( \mathbf{p}_{0},t\right) =\frac{\Delta _{p_{0},\mu ,T}}{2}%
\cos \Omega _{n}t+\frac{\Xi _{p_{0},\mu ,T}}{2}.  \label{sol4}
\end{equation}%
At one-photon interband excitation (when $\chi _{0}<<1$) (\ref{scoup}) one
can omit nonlinear over $E_{0}$ terms and for Rabi frequency we have: 
\begin{equation*}
\Omega _{1}=\frac{eE_{0}\left\vert \sin \Theta \left( \mathbf{p}_{0}\right)
\right\vert }{2p_{0}}.
\end{equation*}%
Taking into account resonant condition $2p_{0}\mathrm{v}_{F}\simeq \hbar
\omega $, the latter can be expressed through the interaction parameter $%
\chi _{0}=eE_{0}\mathrm{v}_{F}/(\hbar \omega ^{2})$: 
\begin{equation*}
\Omega _{1}=\omega \left\vert \sin \Theta \left( \mathbf{p}_{0}\right)
\right\vert \chi _{0}.
\end{equation*}%
\begin{figure}[tbp]
\includegraphics[width=.5\textwidth]{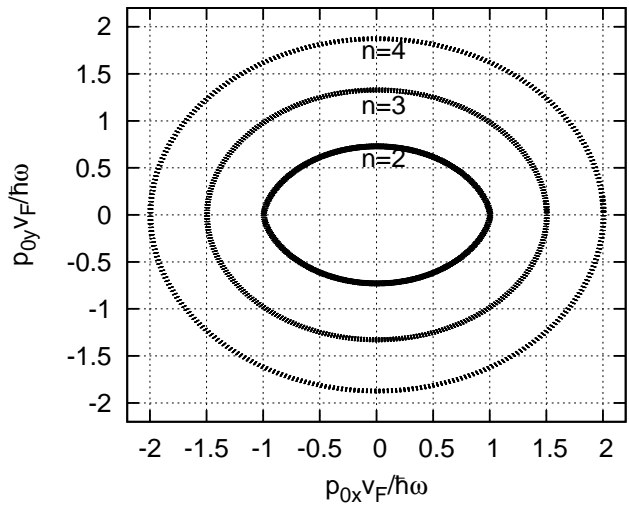}
\caption{Isolines of quasienergy corresponding to resonant condition $2%
\mathcal{E}_{E_{0}}\left( \mathbf{p}_{0}\right) /\hbar \protect\omega %
\approx n$ (\protect\ref{res}) with detuning $\left\vert \protect\delta %
_{n}\right\vert /\protect\omega =0.02$, for $n=2,3,4$. The electric field
dimensionless parameter is taken to be $\protect\chi _{0}=1.0$. The momentum
components are normalized to $\hbar \protect\omega /\mathrm{v}_{F}$.}
\end{figure}
With increasing of pump wave intensity, the Rabi oscillations appear
corresponding to multiphoton transitions. At that one should take into
account the intensity effect of the pump wave on the quasienergy spectrum
(Stark shift due to free-free intraband transitions) and the dynamic Stark
shift due to virtual nonresonant transitions. For $\chi \sim 1$, the
probabilities of multiphoton transitions are essential up to photon numbers $%
n\sim 5$. For this photon numbers the Stark shift (\ref{sol1}) is not
essential, while the modification in the quasienergy spectrum is
considerable. Isolines of quasienergy spectrum, defined by Eq. (\ref{quasi}%
), are no longer circles but ellipse-like. In Fig. 2, isolines of
quasienergy corresponding to resonant condition $2\mathcal{E}_{E_{0}}\left( 
\mathbf{p}_{0}\right) /\hbar \omega \approx n$ (\ref{res}) with detuning $%
\left\vert \delta _{n}\right\vert /\omega =0.02$, for $n=2,3,4$ are shown in
the multiphoton interaction regime: $\chi _{0}=1$. As is seen, modification
of unperturbed energy spectrum in the direction perpendicular to the
electric field vector, is considerable (for example, at $n=3$ we have $13\%$
deviation). Thus, in the multiphoton interaction regime one should expect
photoexcitation of particle distribution function just along the modified
isolines, in accordance with Eq. (\ref{sol4}).

\section{Numerical treatment}

We have also performed numerical simulations and integrated Eqs. (\ref{f1})
and (\ref{f2}) with the fourth-order adaptive Runge-Kutta method. As we
mainly interested by interband multiphoton transitions, for all calculations
the chemical potential and temperature are fixed and are taken to be $\mu
/\hbar \omega =0.1$ and $T/\hbar \omega =5\times 10^{-3}$.

\begin{figure}[tbp]
\includegraphics[width=.5\textwidth]{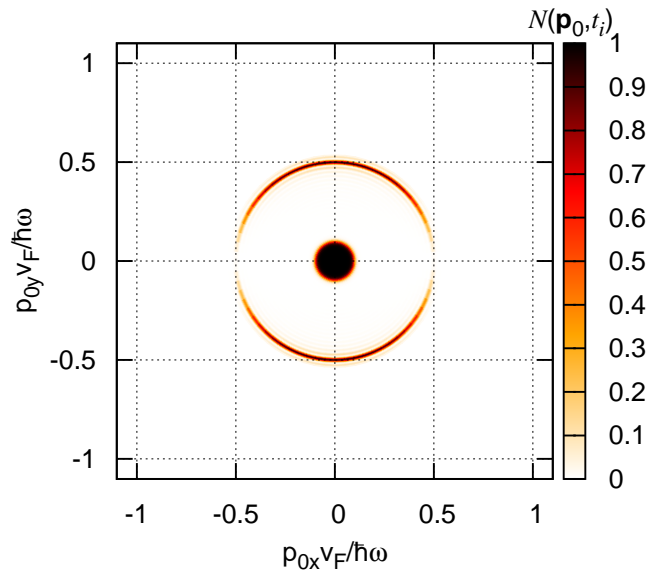}
\caption{(Color online) Creation of particle-hole pair in graphene at the
one-photon resonant excitation. Particle distribution function $\mathcal{N}%
\left( \mathbf{p}_{0},t\right) $ (in arbitrary units) at instant $t_{i}=25%
\mathcal{T}$, as a function of scaled dimensionless momentum components $%
\left\{ p_{0x}\mathrm{v}_{F}/\hbar \protect\omega ,p_{0y}\mathrm{v}%
_{F}/\hbar \protect\omega \right\} $. The electric field dimensionless
parameter is $\protect\chi _{0}=0.02$.}
\end{figure}

\begin{figure}[tbp]
\includegraphics[width=.5\textwidth]{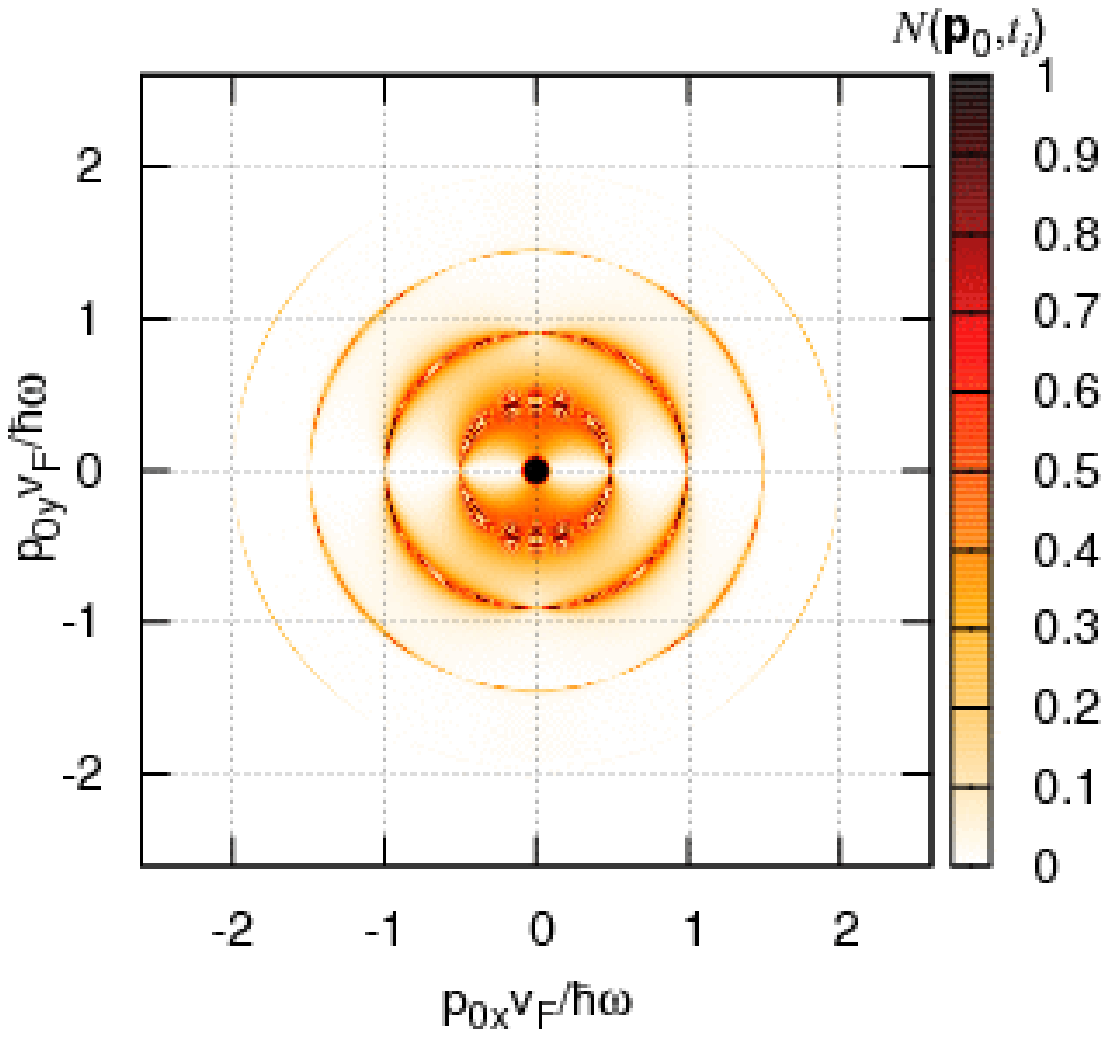}
\caption{(Color online) Creation of particle-hole pair in graphene at the
multiphoton resonant excitation. Particle distribution function $\mathcal{N}%
\left( \mathbf{p}_{0},t\right) $ (in arbitrary units) at instant $t_{i}=100%
\mathcal{T}$, as a function of scaled dimensionless momentum components $%
\left\{ p_{0x}\mathrm{v}_{F}/\hbar \protect\omega ,p_{0y}\mathrm{v}%
_{F}/\hbar \protect\omega \right\} $. The electric field dimensionless
parameter is $\protect\chi _{0}=0.5$.}
\end{figure}

\begin{figure}[tbp]
\includegraphics[width=.5\textwidth]{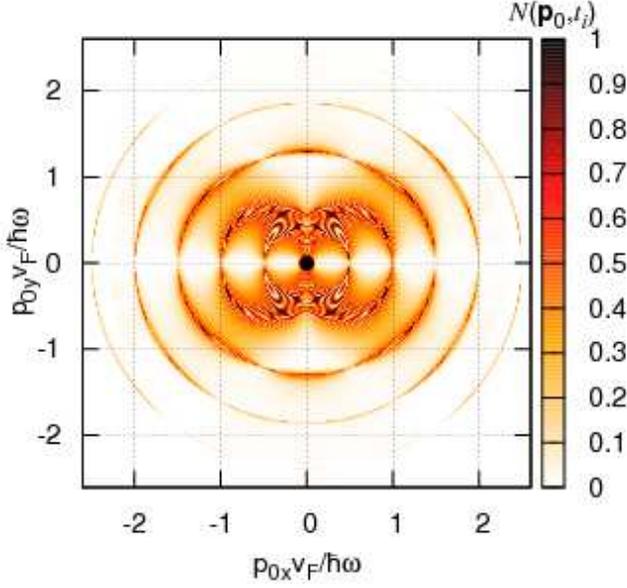}
\caption{(Color online) Same as Fig. 4, but for stronger wave-field with $%
\protect\chi _{0}=1$.}
\end{figure}

\ In Figures 3-5 photoexcitations of Fermi-Dirac sea is presented: density
plot of the particle distribution function $\mathcal{N}%
\left( \mathbf{p}_{0},t\right) $ is
shown for various pump wave intensities and instants. In Fig. 3
corresponding to $\chi _{0}=0.02$, we see only creation of particle-hole
pair in graphene at the one-photon resonant excitation. In Fig. 4, the pump
wave intensity is larger: $\chi _{0}=0.5$ and, as a consequence, we see
resonant rings corresponding to multiphoton excitation up to four photons.
In Fig. 5, which corresponds to\ $\chi _{0}=1$, it is clearly seen also the
ring for five photon resonance. At that, the ring for one-photon excitation
is smeared, since Stark shift for this energy is comparable to $\hbar \omega 
$ and the condition (\ref{condition}) for resonant Rabi oscillations at
one-photon excitation is not fulfilled. As is seen from Figs. 4 and 5, the
excitation of Fermi-Dirac sea takes place along the ellipse-like isolines of
quasienergy spectrum defined by Eq. (\ref{quasi}), in accordance with
analytical treatment (see, Fig. 2). 
\begin{figure}[tbp]
\includegraphics[width=.5\textwidth]{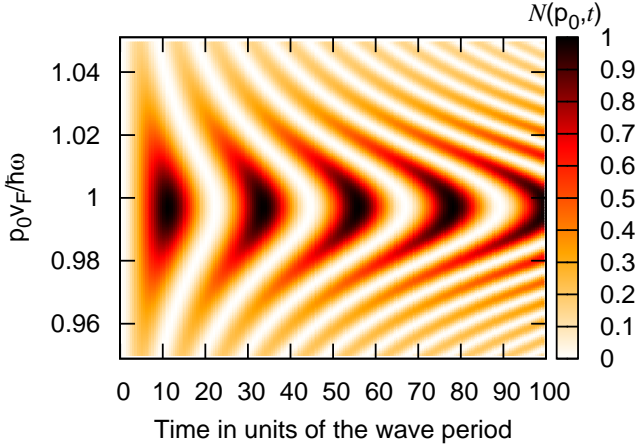}
\caption{(Color online) Two-photon resonance ($n=2$). Rabi oscillations of
the particle distribution function $\mathcal{N}\left( p_{0},t\right) $ for
the fixed angle $\Theta \left( \mathbf{p}_{0}\right) =0.2\ \mathrm{rad}$
versus the scaled dimensionless momentum $p_{0}\mathrm{v}_{F}/\hbar \protect%
\omega $. The electric field dimensionless parameter is $\protect\chi %
_{0}=0.5$.}
\end{figure}
\begin{figure}[tbp]
\includegraphics[width=.5\textwidth]{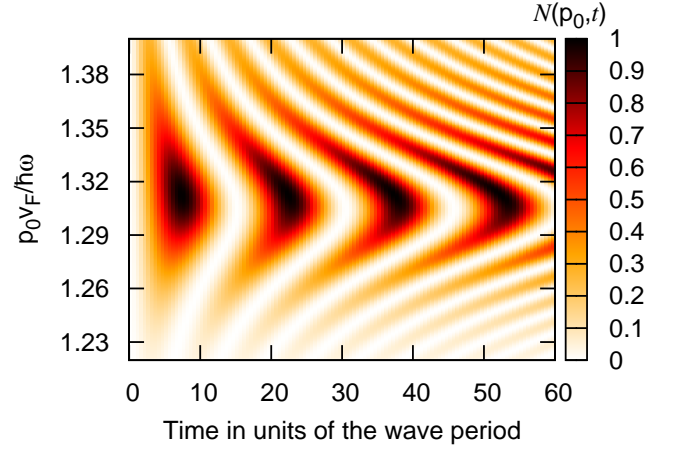}
\caption{(Color online) Same as Fig. 6, but for three-photon resonance ($n=3$%
). The angle is taken to be $\Theta \left( \mathbf{p}_{0}\right) =\protect%
\pi /2\ \mathrm{rad}$\textrm{. }The electric field dimensionless parameter
is $\protect\chi _{0}=1.0$.}
\end{figure}
\begin{figure}[tbp]
\includegraphics[width=.5\textwidth]{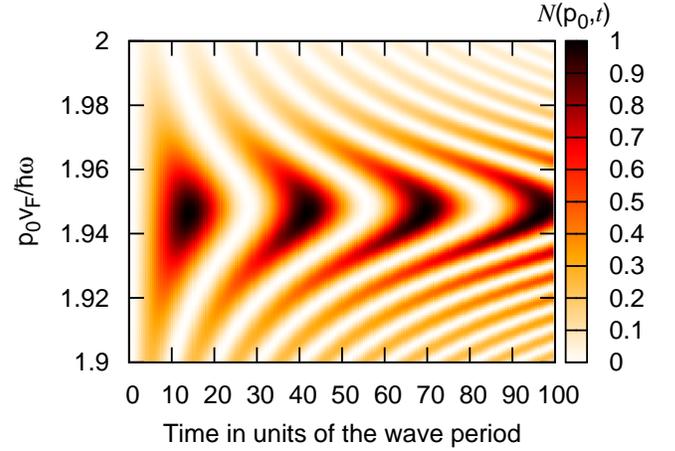}
\caption{(Color online) Four-photon resonance ($n=4$). Same as Fig. 7, but
for the angle $\Theta \left( \mathbf{p}_{0}\right) =0.6\ \mathrm{rad}$. The
electric field dimensionless parameter is $\protect\chi _{0}=1.0$.}
\end{figure}

To show the dynamics of multiphoton excitation of Fermi-Dirac sea in Figs.
6-8, we present Rabi oscillations of the particle distribution function $%
\mathcal{N}\left( p_{0},t\right) $ for the fixed angles versus the scaled
dimensionless momentum $p_{0}\mathrm{v}_{F}/\hbar \omega $. Figure 6
corresponds to two-photon resonance for the angle $\Theta \left( \mathbf{p}%
_{0}\right) =0.2\ \mathrm{rad}$ and $\chi _{0}=0.5$. It is clearly seen Rabi
oscillations of $\mathcal{N}\left( p_{0},t\right) $ with the mean period $%
\mathcal{T}_{R}=22\mathcal{T}$ ($\mathcal{T}$ is the wave period). Rabi
oscillations for three-photon resonance is displayed in Fig. 7 for the angle 
$\Theta \left( \mathbf{p}_{0}\right) =\pi /2\ \mathrm{rad}$\textrm{\ }and%
\textrm{\ }$\chi _{0}=1$. Here mean Rabi period is $\mathcal{T}_{R}=15%
\mathcal{T}$. Four-photon resonant Rabi oscillations with the mean period $%
\mathcal{T}_{R}=28\mathcal{T}$, at the angle $\Theta \left( \mathbf{p}%
_{0}\right) =0.6\ \mathrm{rad}$ is shown in Fig. 8. 
\begin{figure}[tbp]
\includegraphics[width=.5\textwidth]{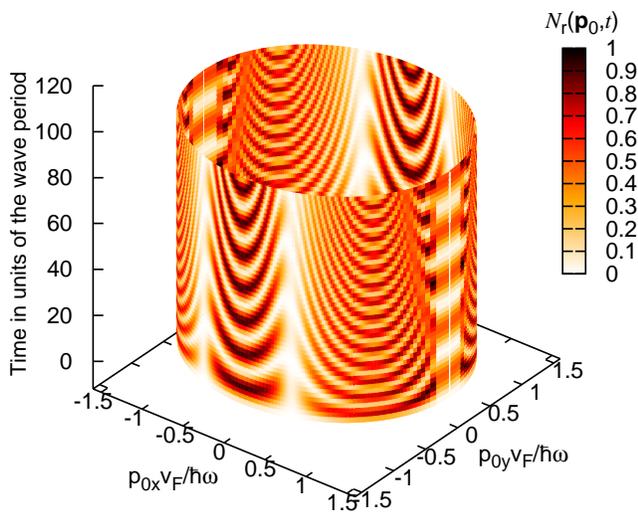}
\caption{(Color online) Colored 4D plot of Rabi oscillations of the particle
distribution function $\mathcal{N}_{r}\left( \mathbf{p}_{0},t\right) $ for
three-photon resonance on isosurface $2\mathcal{E}_{E_{0}}\left( \mathbf{p}%
_{0}\right) /\hbar \protect\omega =3$. The electric field dimensionless
parameter is $\protect\chi _{0}=1.0$.}
\end{figure}

\begin{figure}[tbp]
\includegraphics[width=.5\textwidth]{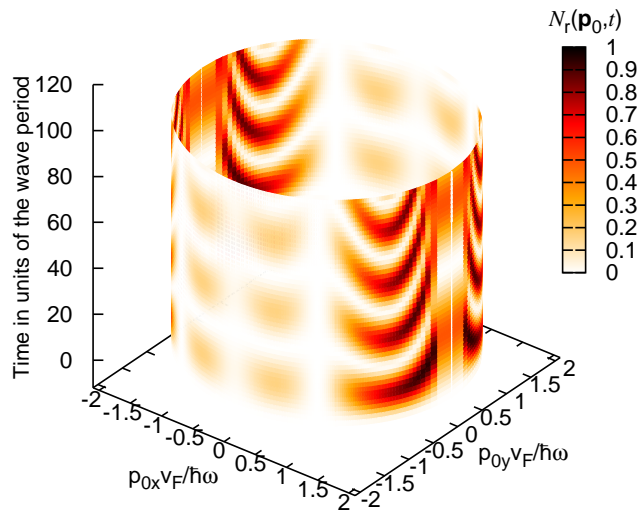}
\caption{(Color online) Same as Fig. 9, but for four-photon resonance: $2%
\mathcal{E}_{E_{0}}\left( \mathbf{p}_{0}\right) /\hbar \protect\omega =4$.}
\end{figure}

To show the dependence of Rabi oscillations on the angle $\Theta \left( 
\mathbf{p}_{0}\right) $, in Figs. 9 and 10 colored 4D density plot of Rabi
oscillations of the particle resonant distribution function $\mathcal{N}%
_{r}\left( \mathbf{p}_{0},t\right) $ on isosurfaces $2\mathcal{E}%
_{E_{0}}\left( \mathbf{p}_{0}\right) /\hbar \omega =n$, for $n=3,4$ are
shown at $\chi _{0}=1$. As is seen from these figures, for odd photon
resonance the excited distribution function $\mathcal{N}_{r}\left( \mathbf{p}%
_{0},t\right) $ is maximal at the angle $\Theta \left( \mathbf{p}_{0}\right)
=\pi /2\ \mathrm{rad}$ (perpendicular to applied electric field vector),
while for even photon resonance at $\Theta \left( \mathbf{p}_{0}\right) =\pi
/2\ \mathrm{rad}$ we have a minimum and main excitation takes place close to 
$\Theta \left( \mathbf{p}_{0}\right) =\pi /4$. The latter is connected with
the fact that coupling term in Eqs. (\ref{ff1}), (\ref{ff2}) at $\Theta
\left( \mathbf{p}_{0}\right) =\pi /2\ \mathrm{rad}$ contains only odd
harmonics of the pump wave: $G_{2s}\left( 0,p_{0y},E_{0}\right) =0$.

Summarizing, we see that numerical simulations are in agreement with
analytical treatment in multiphoton resonant approximation and confirm
simple physical picture described in previous section. We see that with
laser fields of moderate intensities one can observe multiphoton excitation
of Fermi-Dirac sea, the probabilities of which are comparable to and
sometimes larger than one-photon excitation probability.

\section{Conclusion}

We have presented a theoretical treatment of the coherent nonlinear response
of a graphene under multiphoton interband excitation by a laser radiation.
The evolutionary equation for single-particle density matrix is formulated
arising from the second quantized formalism. The time-dependent
single-particle density matrix in the given field of a laser radiation is
calculated in the multiphoton resonant approximation. The Rabi oscillations
of Fermi-Dirac sea at multiphoton excitation depending on the time,
momentum, and photon number have been considered and analyzed also on the
base of numerical simulations. The results obtained demonstrate that the
Rabi oscillations of Fermi-Dirac sea corresponding to multiphoton excitation
can already be observed for such laser fields where the work of electric
field on the wave period is comparable to photon energy $\varepsilon
_{\gamma }=\hbar \omega $. For mid-infrared lasers $\varepsilon _{\gamma
}\sim 0.1$ $\mathrm{eV}$, multiphoton interaction regime can be achieved at
the intensities $I>10^{7}\ \mathrm{W\ cm}^{-2}$, for the time scales $1.0\ 
\mathrm{ps}$\textrm{. }For near-infrared range of frequencies\textrm{\ }$%
\varepsilon _{\gamma }\sim 1.$ $\mathrm{eV}$,\textrm{\ }multiphoton
interaction regime can be achieved at the intensities $I>10^{11}\ \mathrm{W\
cm}^{-2}$,\textrm{\ }for the time scales $100.0\ \mathrm{fs}$\textrm{.}

\begin{acknowledgments}
This work was supported by State Committee of Science (SCS) of Republic
of Armenia (RA), Project No. 11RB-006.

\end{acknowledgments}


\begin{thebibliography}{99}
\bibitem{Nov1} K. S. Novoselov, A. K. Geim, S. V. Morozov, D. Jiang, Y.
Zhang, S. V. Dubonos, I. V. Grigorieva, and A. A. Firsov, Science \textbf{306%
}, 666 (2004).

\bibitem{Nov2} A. H. Castro Neto, F. Guinea, N. M. R. Peres, K. S.
Novoselov, and A. K. Geim, Rev. Mod. Phys. \textbf{81}, 109 (2009).

\bibitem{Gaim} A. K. Geim, Science \textbf{324}, 1530 (2009).

\bibitem{Nov3} K.S. Novoselov, A.\ K. Geim, S. V. Morozov, D. Jiang, M. I.
Katsnelson, I. V. Grigorieva, S.V. Dubonos and A. A. Firsov, Nature, \textbf{%
438}, 197 (2005).

\bibitem{GQED1} G. W. Semenoff, Phys. Rev. Lett. \textbf{53}, 2449 (1984).

\bibitem{GQED2} F. D. M. Haldane, Phys. Rev. Lett. \textbf{61}, 2015 (1988).

\bibitem{GQED4} M. I. Katsnelson, K. S. Novoselov, and A. K. Geim, Nature
Phys. \textbf{2}, 620 (2006).

\bibitem{GQED5} V. V. Cheianov, V. I. Fal'ko, and B. L. Altshuler, Science 
\textbf{315}, 1252 (2007).

\bibitem{GQED6} M. I. Katsnelson and K. S. Novoselov, Solid State Commun. 
\textbf{143}, 3 (2007).

\bibitem{Book} H.K. Avetissian, \textit{Relativistic Nonlinear
Electrodynamics} (Springer, New York, 2006).

\bibitem{K1} M.I. Katsnelson, Materials Today \textbf{10}, 20 (2007).

\bibitem{K2} V. V. Cheianov and V. I. Fal'ko, Phys. Rev. B \textbf{74},
041403 (2006).

\bibitem{K3} C. W. J. Beenakker, Rev. Mod. Phys. \textbf{80}, 1337 (2008).

\bibitem{Z1} C. Itzykson, J. -B. Zubar,\textit{\ Quantum Field Theory}
(Dover, 2006).

\bibitem{Z2} M. I. Katsnelson, Eur. Phys. J. B \textbf{51}, 157 (2006).

\bibitem{Z3} J. Cserti and Gy. D\'{a}vid, Phys. Rev. B \textbf{74}, 172305
(2006).

\bibitem{Z4} J. Schliemann, New J. Phys. \textbf{10}, 043024 (2008).

\bibitem{S1} J. Schwinger, Phys. Rev. \textbf{82}, 664 (1951).

\bibitem{S2} D. Allor, T. D. Cohen, D. A. McGady, Phys.Rev. D \textbf{78},
096009 (2008).

\bibitem{S3} B. D\'{o}ra and R. Moessner, Phys. Rev. B \textbf{81}, 165431
(2010).

\bibitem{cn1} A.W. W. Ludwig, M. P. A. Fisher, R. Shankar, and G. Grinstein,
Phys. Rev. B \textbf{50}, 7526 (1994).

\bibitem{cn2} K. Ziegler, Phys. Rev. B \textbf{75}, 233407 (2007).

\bibitem{pair} H. K. Avetissian, A. K. Avetissian, G. F. Mkrtchian, Kh. V.
Sedrakian, Phys. Rev. E \textbf{66},\textbf{\ }016502 (2002).

\bibitem{Eb-Sh} L. Allen, J. H. Eberly, \textit{Optical Resonance and Two
Level Atoms} (Wiley-Interscience, New York, 1975); B.W. Shore, \textit{%
Theory of Coherent Atomic Excitation} (Wiley-Interscience, New York, 1990);
M. O. Scully and M. S. Zubairy, \textit{Quantum Optics} (Cambridge
University Press, Cambridge, U.K., 1997).

\bibitem{Rs1} F. Rossi and T. Kuhn, Rev. of Mod. Phys., 74, 895 (2002); V.
M. Axt and T. Kuhn, Rep. Prog. Phys. 67, 433 (2004).

\bibitem{Rabi1} E. G. Mishchenko, Phys. Rev. Lett. \textbf{103}, 246802
(2009).

\bibitem{Rabi2} P.N. Romanets and F.T. Vasko, Phys. Rev. B \textbf{81},
241411(R) (2010).

\bibitem{Rabi3} B. D\'{o}ra, K. Ziegler, P. Thalmeier, and M. Nakamura,
Phys. Rev. Lett. \textbf{102}, 036803 (2009).

\bibitem{MRE1} H. K. Avetissian, G. F. Mkrtchian, Phys. Rev. A \textbf{66},
033403 (2002).

\bibitem{MRE2} H. K. Avetissian, B. R. Avchyan, and G. F. Mkrtchian, Phys.
Rev. A \textbf{77}, 023409 (2008).

\bibitem{MRE3} H. K. Avetissian, B. R. Avchyan, G. F. Mkrtchian, Phys. Rev.
A \textbf{82}, 063412 (2010).
\end{thebibliography}
\end{document}